\documentclass[a4paper,11pt]{article}
\usepackage{pos}

\usepackage[font=small,labelfont=bf,justification=centerlast]{caption}
\usepackage{array,multirow,makecell}
\usepackage{subcaption}
\usepackage{slashed}
\usepackage{hypernat}

\usepackage{graphicx}
\usepackage[nointegrals]{wasysym}
\usepackage{rotating}
\usepackage{color}
\usepackage{epsfig}
\usepackage{bm}
\usepackage{hyperref}
\usepackage{url}
\usepackage{slashed}

\usepackage{float}

\newcommand{\be}{\begin{equation}}
\newcommand{\ee}{\end{equation}}
\newcommand{\bea}{\begin{eqnarray}}
\newcommand{\eea}{\end{eqnarray}}

\def\lg{{\mathchoice{~\raise.58ex\hbox{$<$}\mkern-14.8mu\lower.52ex\hbox{$>$}~}
                    {~\raise.58ex\hbox{$<$}\mkern-14.8mu\lower.52ex\hbox{$>$}~}
                    {\raise.59ex\hbox{{$\scriptscriptstyle <$}}\mkern-12.8mu%
                     \lower.01ex\hbox{{$\scriptscriptstyle >$}}}   {}   }}
\def\gl{{\mathchoice{~\raise.58ex\hbox{$>$}\mkern-12.8mu\lower.52ex\hbox{$<$}~}
                    {~\raise.58ex\hbox{$>$}\mkern-12.8mu\lower.52ex\hbox{$<$}~}
                    {\raise.62ex\hbox{{$\scriptscriptstyle >$}}\mkern-12.0mu%
                     \lower.05ex\hbox{{$\scriptscriptstyle <$}}}  {}    }}

\title{Self-interacting dark matter from late decays and the $H_0$ tension}
\ShortTitle{SIDM from late decays and the $H_0$ tension}

\author[*,\dag]{Krzysztof~Jod\l{}owski}
\notes{\note{K. J. is supported in part by the National Science Centre, Poland, research Grant No. 2015/ A/ST2/00748.}\note{Content of this proceedings is based on previous work with Andrzej Hryczuk \cite{Hryczuk:2020jhi}.}}

\affiliation{National Centre for Nuclear Research,\\
 Pasteura 7, 02-093 Warsaw, Poland}

\emailAdd{krzysztof.jodlowski@ncbj.gov.pl}

\abstract{
We investigate a mechanism for the production of self-interacting dark matter based on WIMP-like messenger state decays into dark matter and dark radiation occurring after recombination. Such a transition leads to a mild relaxation of the Hubble tension, and at the same time may resolve the small-scale structure problems of the $\Lambda$CDM. We illustrate this mechanism in the Higgs portal dark matter model, which we find to be a promising route.
}

\FullConference{%
The European Physical Society Conference on High Energy Physics (EPS-HEP2021),\\
26-30 July 2021 \\
Online conference, jointly organized by Universität Hamburg and the research center DESY
}

\begin{document}
\maketitle

\section{Introduction}

The standard model of cosmology incorporates dark matter (DM) as a cold noninteracting matter component with a constant equation of state which leads to remarkably successful description of the Universe \cite{Planck:2018vyg}, especially at large scales.
However, at smaller scales severe discrepancies between theoretical predictions based on such assumptions and observations were observed \cite{Bullock:2017xww}. Moreover, recent advances in observational cosmology identified another source of tension within the $\Lambda$CDM, most notably the values of the Hubble rate parameter, $H_0$, \cite{Verde:2019ivm} and to lesser extend the `clumpiness' parameter, $S_8=\sigma_8\left(\Omega_{\mathrm{matter}}/0.3\right)^{0.5}$.

\section{The mechanism}
\label{sec:model}

It is well-known that self-interactions within the dark sector (DS) provide a better fit to galactic-scale data than $\Lambda$CDM \cite{Tulin:2017ara}. Moreover, transforming a small fraction of DM into radiation has been shown \cite{Blackadder:2015uta} to significantly reduce the $H_0$ tension - but also note recent review of $H_0$-tension solutions \cite{Schoneberg:2021qvd} for updated analysis. It is interesting to ponder whether modification of only the DM component can provide a solution to both small and large-scale problems of $\Lambda$CDM. In light of this, we propose a  mechanism of self-interacting DM (SIDM) production taking place through late decays of a messenger WIMP-like state mainly into SIDM, but also sub-dominantly into dark radiation.

\begin{figure}[h]
\centering
\includegraphics[scale=0.205]{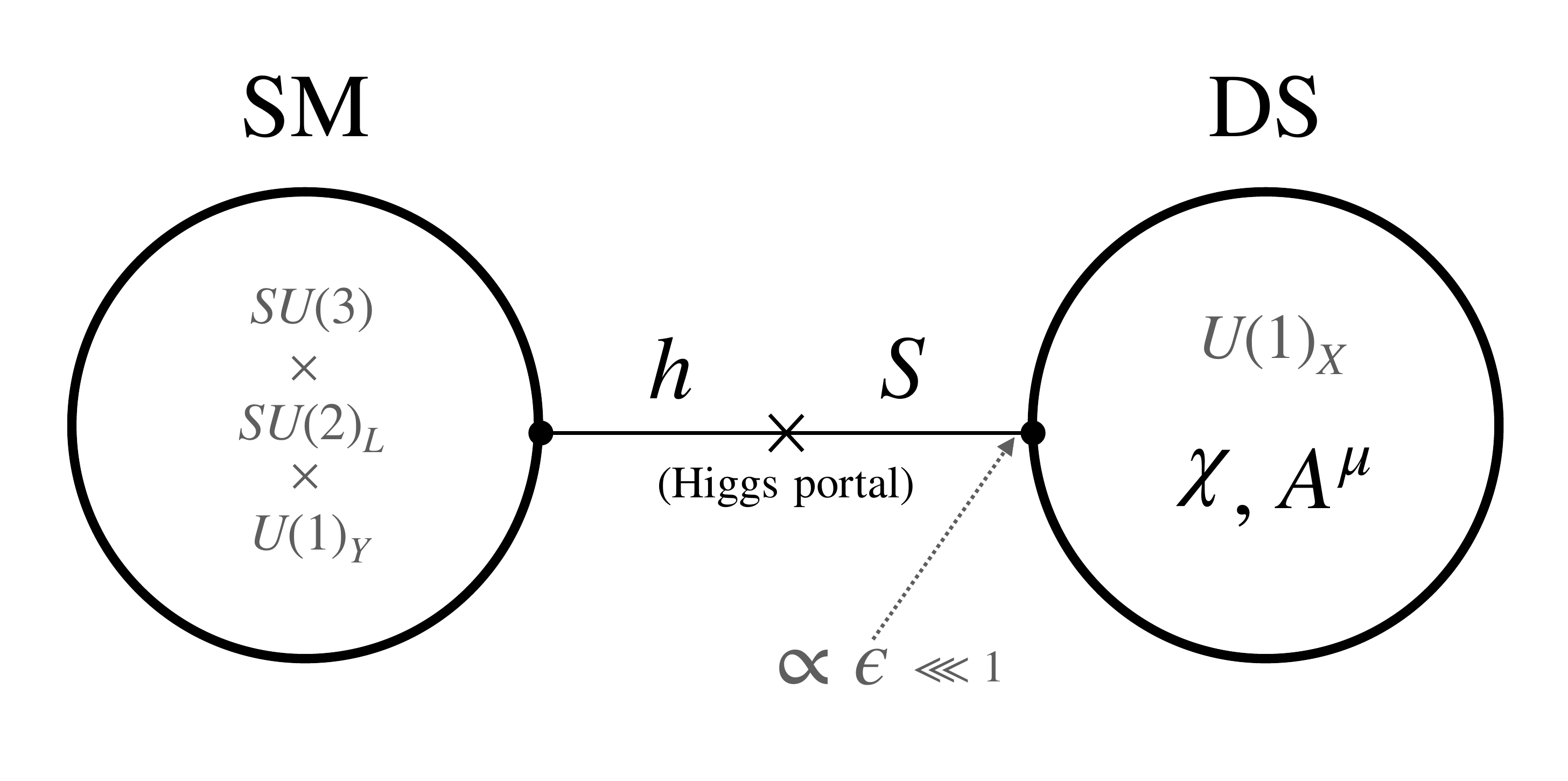}
\caption{Setup where the proposed mechanism is realized. SM is connected through a Higgs portal connector $S$ to the DS, built from a Dirac fermion $\chi$ charged under $U(1)_{\textrm{dark}}$ with a massive gauge field $A^\mu$.}
\label{fig:model}
\end{figure}

\begin{figure*}[h]
\begin{subfigure}{1\textwidth}
\centering
\includegraphics[width=0.79\textwidth]{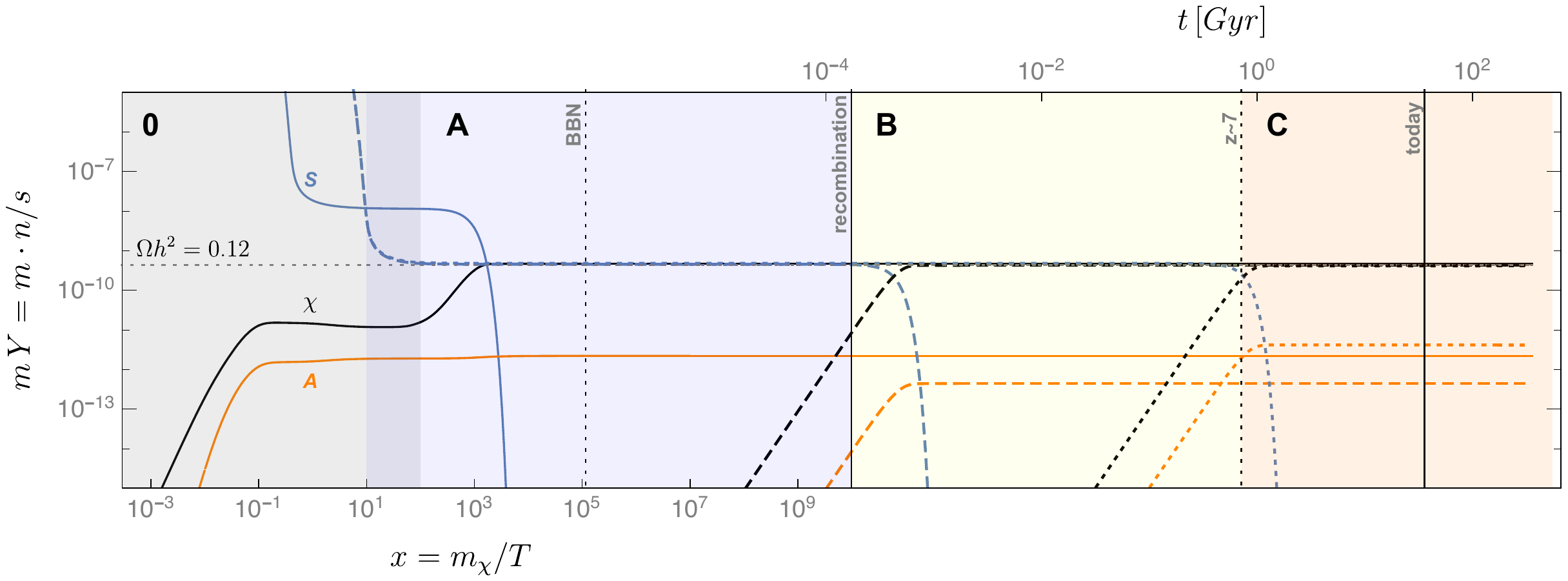}
\end{subfigure}
\caption{\textbf{Thermal history}: evolution of energy densities of $S$ (blue), $\chi$ (black) and $A^\mu$ (orange) as a function of $x=m_\chi/T$ for parameters which lead to early (regime A, solid lines: very weak $\lesssim\epsilon \lesssim$ weak - SIDM viable), late (regime B, dashed: ultra weak $\lesssim \epsilon \lesssim$ very weak - SIDM with an impact on the $H_0$ tension) and very late (regime C, dotted: $\epsilon \lesssim$ ultra weak - two component DM (dominant pseudo-WIMP $S$ and small component of ultrastrong self-interacting dark matter \cite{Pollack:2014rja}) with an impact on the $H_0$ tension) decays of $S$.}
\label{fig:thermal_history}
\end{figure*}

\subsection{The SM-DS coupling through a portal }
Going forward, we concentrate on a Higgs portal scenario, which is illustrated in Fig.~\ref{fig:model}.
The DS after the $U(1)_{\textrm{dark}}$ breaking is described by the following Lagrangian:
\begin{eqnarray}
\mathcal{L}^{\rm DS} =\ & &
\bar{\chi}( i \gamma_\mu \partial^\mu-m_{\chi})\chi + \frac{1}{2}m_A^2 A_\mu A^\mu + igA^\mu \bar\chi \gamma_\mu \chi + \epsilon \, S \bar{\chi} \chi,
\label{eq:LDS}
\end{eqnarray}
while the connection with the DM is given by the portal interactions:
\begin{eqnarray}
\mathcal{L}^{\rm portal} \supseteq\ && \epsilon\,\mu_{HS} S \, H^{\dagger} H    +   \lambda_{HS} \, S^2 H^{\dagger} H \, ,
\label{eq:Lportal}
\end{eqnarray}
where $H$ denotes the SM Higgs boson doublet. It is natural to assume that $S$ is a pseudo-WIMP - particle freezing-out from thermal plasma that is almost stable due to a broken symmetry (in our case, $Z_2$: $S\to -S$). One can parameterize the symmetry breaking by a small parameter $\epsilon$ which value leads into four distinctive regimes of thermal history and impact on $H_0$ as illustrated in Fig.~\ref{fig:thermal_history}

\section{Fit to cosmological scan}

\begin{figure}[!t]
\centering
\hspace*{-0.6cm}\includegraphics[scale=0.35]{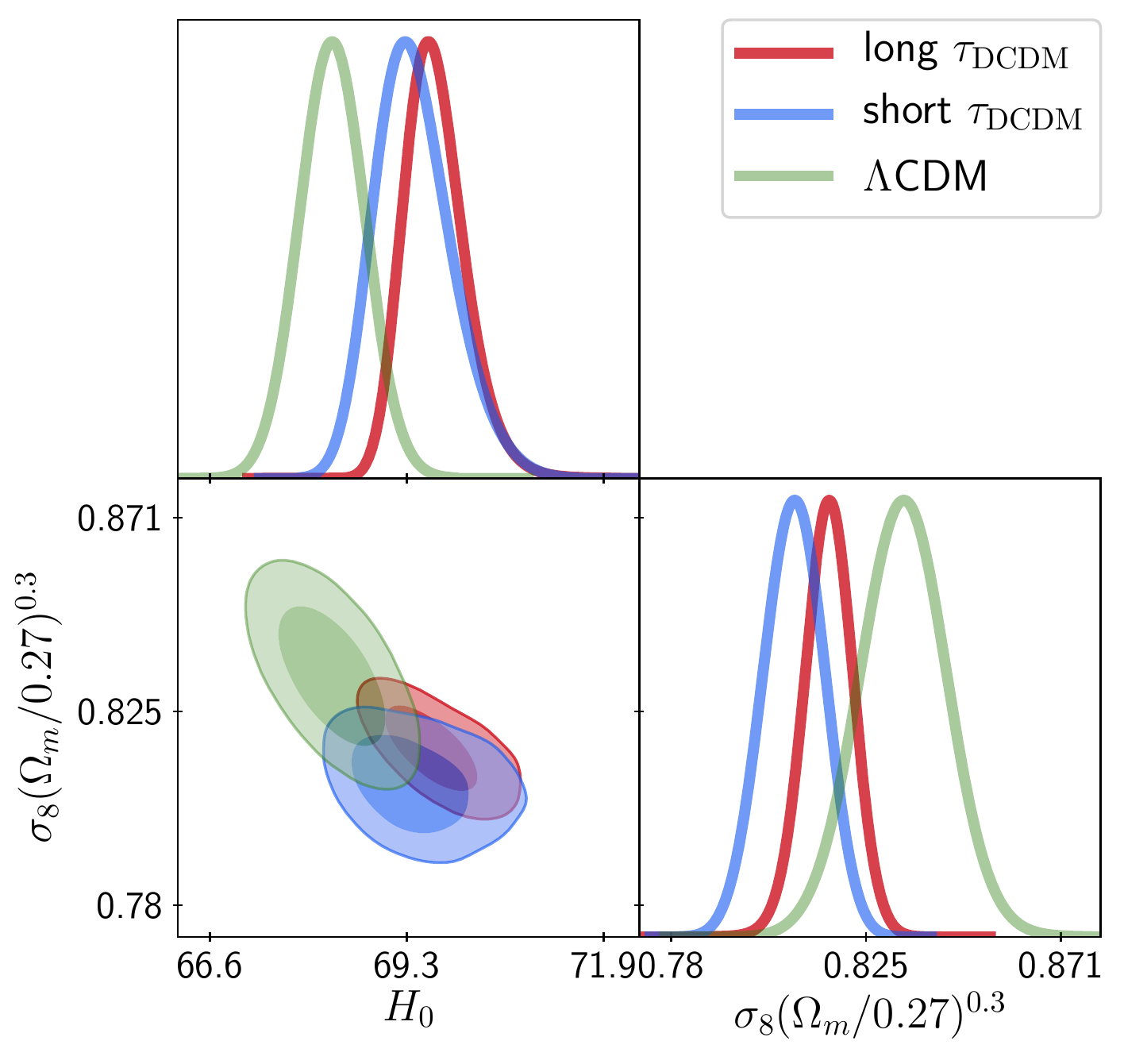}
\hspace*{0.5cm}\includegraphics[scale=0.19]{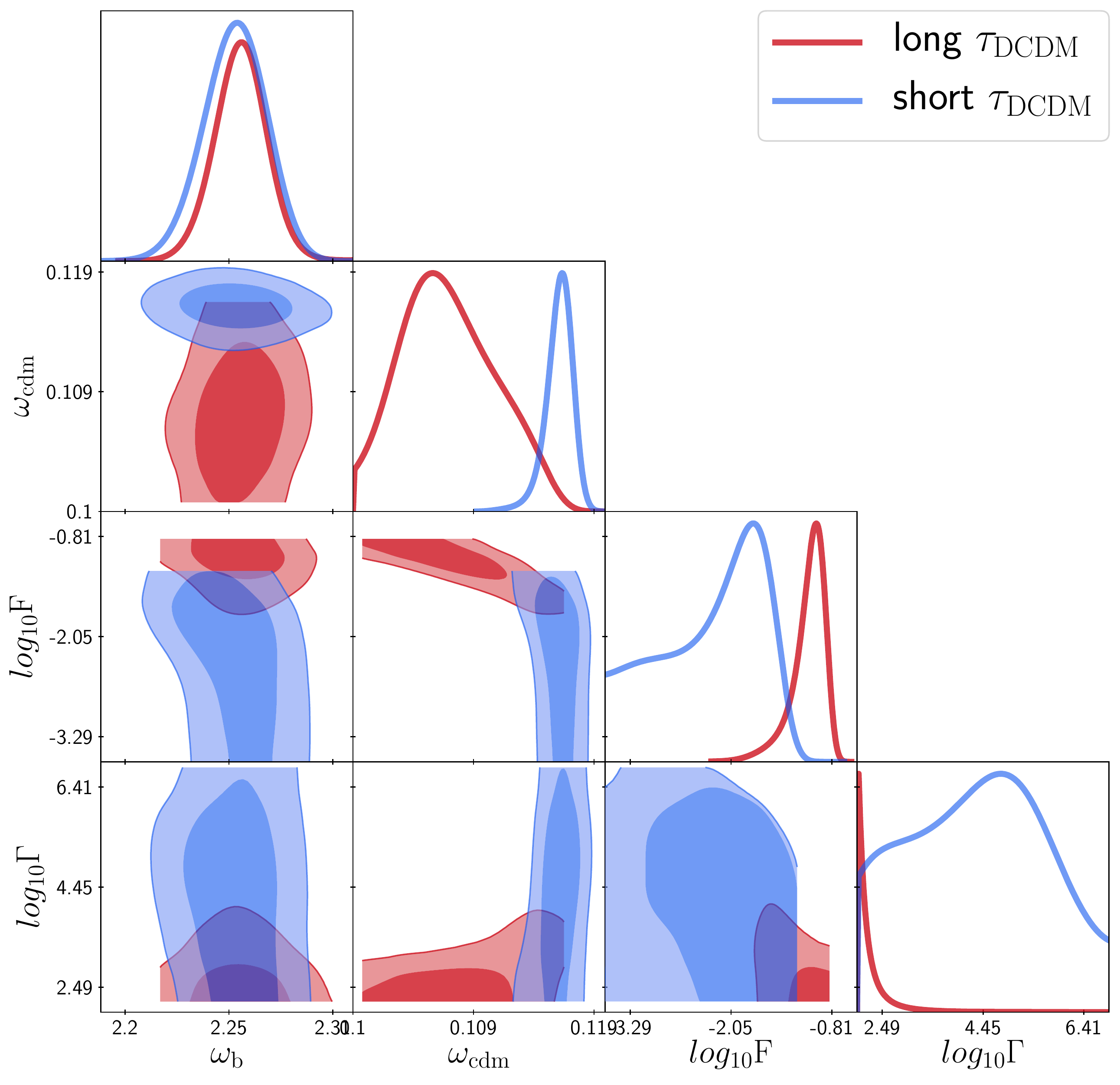}
\caption{Constraints on the cosmological parameters for the regimes B (blue) and C (red) of the DCDM and $\Lambda$CDM. Both regimes of DCDM lead to mildly better values of $H_0$ and $\sigma_8\left(\Omega_{\mathrm{matter}}/0.27\right)^{0.3}$ over $\Lambda$CDM.
\label{fig:cosmofit}}
\end{figure}

To determine the impact of decaying DM (DCDM) on cosmological evolution, we performed Monte Carlo Markov Chain scan using \texttt{CLASS} together with \texttt{MontePython}~\cite{Brinckmann:2018cvx}. We used datasets from both early and late Universe observations. The results are presented in Fig.~\ref{fig:cosmofit} and show 1-$\sigma$ preferred regions for regime B (blue) and C (red) where the $H_0$ tension is mildly alleviated.

\subsection{The SIDM from late decays regime and the uSIDM regime}

\begin{figure}[!t]
\centering
\hspace*{-0.6cm}\includegraphics[scale=0.39]{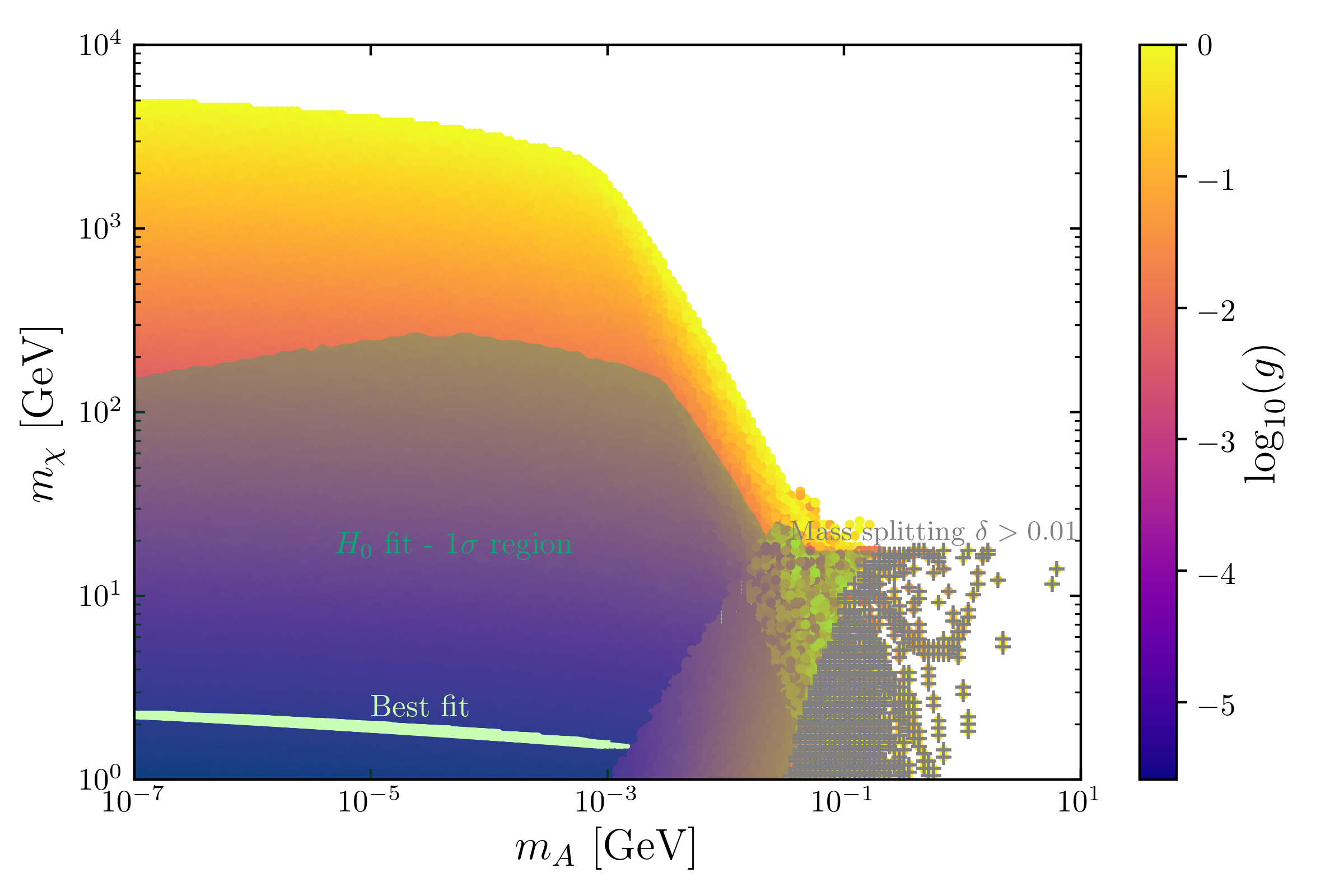}
\hspace*{0.5cm}\includegraphics[scale=0.38]{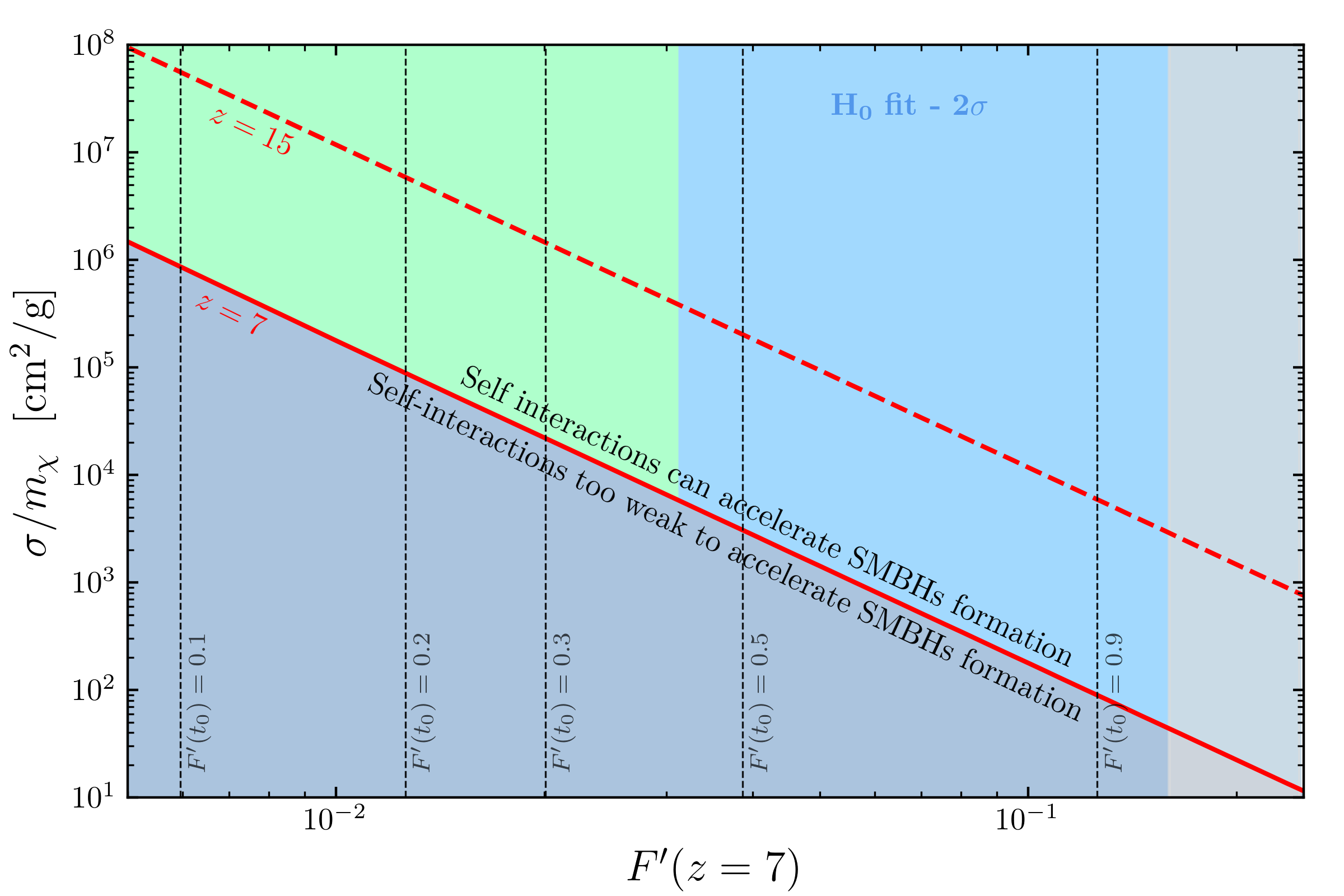}
\setlength{\belowcaptionskip}{-15pt}
\caption{Cosmological scan results for the SIDM regime B (left) and uSIDM regime C (right). On the left, color coding indicates the value of the coupling $g$ leading to $\sigma/m_\chi \sim (1\pm10\%)$ cm$^2$/g.
On the right, self-interactions strong enough to accelerate SMBHs formation rates are denoted as the blue and green areas.
\label{fig:cosmofit_compare_cosmofit_short}}
\end{figure}

For very small values of $\epsilon$, $S$ decays after the recombination which modifies the evolution of background and perturbation quantities over $\Lambda$CDM while still producing SIDM. The results for this regime are presented on the left of Fig.~\ref{fig:cosmofit_compare_cosmofit_short}, while on the right one can see the results for even longer $\tau_S\sim 10$ Gyr, where we obtain two-component, ultrastrong self-interacting dark matter.

\section{Conclusions}
\label{sec:conclusions}

Summing up, we studied the cosmological implications of the SIDM production mechanism based on decays of WIMP-like state taking place at a late time. The dominant decay mode is the SIDM production taking place via a tree-level process while higher order processes transfer a small part of the WIMP-like state energy into dark radiation. Such higher order processes naturally lead to transfer of only a small part of DM energy into radiation, which is mandatory due to the stringent CMB bounds, while at the same time improving the fit to the $H_0$ and $S_8$ parameters.

\bibliographystyle{jhep}
\bibliography{biblio}

\end{document}